# Gell-Mann–Low Function in QED

## I. M. Suslov

*Kapitza Institute for Physical Problems, Russian Academy of Sciences, ul. Kosygina 2, Moscow, 117973 Russia*
*e-mail: suslov@kapitza.ras.ru*

The Gell-Mann–Low function $\beta(g)$ in QED ($g$ is the fine structure constant) is reconstructed. At large $g$, it behaves as $\beta_\infty g^\alpha$ with $\alpha \approx 1$ and $\beta_\infty \approx 1$.

Recently a present author [1, 2] developed a method of summation divergent perturbation series with arbitrary coupling constants. With this method, information about all terms of the series is obtained by the interpolation of the known first terms with their Lipatov asymptotics [3]. In this paper, this method will be used to reconstruct the Gell-Mann–Low function in QED.

Lipatov's method [3] is based on the saddle-point calculation of path integrals near instanton configurations and is being questioned because of the possible renormalon contributions [4]. Formally, the asymptotic behavior of perturbation theory is determined by the singularity nearest to the origin in the Borel plane. Whereas the presence of instanton singularities is beyond question, the existence of renormalon singularities has never been proved, which is acknowledged by the most active advocates of this direction [5]. Having been proved in [6], the absence of renormalon singularities in the $\varphi^4$ theory casts doubt on the renormalon concept as a whole, although similar proofs are lacking for other field theories. In such a situation, it is possible to assume that the renormalon singularities are absent.

**1.** The asymptotics of perturbation theory for QED was discussed in the late 1970s [7–9]; all fundamental problems were solved by Bogomolny and Fateyev [8, 9], but no results for specific quantities was obtained. Below, we partially fill this gap.

The vertex with $M$ photon and $2L$ electron free lines is determined by the path integral

$$Z_{M,L} = \int DAD\bar{\psi}D\psi A(x_1)\ldots A(x_M)\psi(y_1)\bar{\psi}(z_1)\ldots\psi(y_L)$$

$$\times \bar{\psi}(z_L)\exp\left\{-\int d^4x\left[\frac{1}{4}(\partial_\mu A_\nu - \partial_\nu A_\mu)^2\right.\right. \quad (1)$$

$$\left.\left.+ \bar{\psi}(i\gamma_\nu\partial_\nu - m + e\gamma_\nu A_\nu)\psi\right]\right\}.$$

Integration with respect to the fermion fields gives

$$Z_{M,L} = \int DA A(x_1)\ldots A(x_M)G(y_1,z_1)\ldots G(y_L,z_L)$$

$$\times \det(i\gamma_\nu\partial_\nu - m + e\gamma_\nu A_\nu) \quad (2)$$

$$\times \exp\left\{-\frac{1}{4}\int d^4x(\partial_\mu A_\nu - \partial_\nu A_\mu)^2\right\} + \ldots,$$

where $G(x, x')$ is the Green's function for the Dirac operator

$$(i\gamma_\nu\partial_\nu - m + e\gamma_\nu A_\nu)G(x, x') = \delta(x - x'), \quad (3)$$

and the ellipsis stands for the terms with other pairings of $\psi(y_i)$ and $\bar{\psi}(z_k)$. Estimations show that the quantity $eA_\nu(x)$ is large for the saddle-point configuration and the asymptotic form of the determinant at $e \longrightarrow i\infty$ can be used, because the growth rate is maximal at imaginary $e$ values [9]:

$$\ln\det(i\gamma_\nu\partial_\nu - m + e\gamma_\nu A_\nu) = \frac{e^4}{12\pi^2}\int d^4x(A_\nu^2)^2. \quad (4)$$

This result is not gauge invariant and is only valid for a specifically chosen gauge; it can be obtained for slowly varying fields or for configurations with a sufficiently high symmetry [9]. Taking Eq. (4) into account, a path integral with effective action

$$S_{\text{eff}}\{A\} = \int d^4x\left\{\frac{1}{4}(\partial_\mu A_\nu - \partial_\nu A_\mu)^2 - \frac{4}{3}g^2(A_\nu^2)^2\right\},$$

$$g = \frac{e^2}{4\pi}, \quad (5)$$

appears in Eq. (2); the asymptotic form of perturbation theory for this action can be found by Lipatov's method. Its structure is determined by the homogeneity properties of the action [10]; when $g^2$ is used as a coupling constant, these properties are the same as in the $\varphi^4$ theory, and the general asymptotic term has the form $cS_0^{-N}\Gamma(N+b)g^{2N}$, where $S_0$ is the instanton action. In actuality, the expansion is in arbitrary integer (not only even) $g$ powers, and the general term is $cS_0^{-N/2}\Gamma(N/2+b)g^N$.[1] Taking the value of instanton action into account, one obtains for the $N$th-order contribution to the vacuum integral ($M = 0$, $L = 0$) [8]:

$$Z_N(-g)^N = \text{const}\left(\frac{3^{3/2}}{4\pi^3}\right)^{N/2}\Gamma\left(\frac{N+r}{2}\right)(-g)^N, \quad (6)$$

where $r = 11$ is the number of zero modes including four translations, a scale transformation, and six four-dimensional rotations (instanton corresponds in symmetry to a rigid body of an irregular shape).

In the general case, the functional form of the result can be found by structural calculations described in [10] and reduced to dimensional analysis. It is easy to show that $e_c \sim N^{-1/4}$ and $A_c(x) \sim N^{1/2}$ for the saddle-point configuration. To find the dimension of $G(x, x')$, consider the Dyson equation

$$G(x, x') = G_0(x - x') - \int d^4 y G_0(x - y)e\gamma_\nu A_\nu(y)G(y, x') \quad (7)$$

which follows from Eq. (3). In order to clarify the structure of the solution, let us consider the scalar analogue of Eq. (7) and assume that the function $A_\nu(x)$ is strongly localized near $x = 0$; one can then set $G(y, x') \approx G(0, x')$ in the integral, after which the equation is easily solved:

$$G(x, x') = G_0(x - x') - \frac{G_0(-x')\int d^4 y G_0(x - y)e\gamma_\nu A_\nu(y)}{1 + \int d^4 y G_0(-y)e\gamma_\nu A_\nu(y)}. \quad (8)$$

Because $eA_\nu(x) \sim N^{1/4}$ and Eq. (8) is finite in the limit $e \longrightarrow \infty$, one has $G(x, x') \sim N^0$. It is natural to expect that this result is general and is not caused by the above assumptions. The $N$th-order contribution to the integral in Eq. (1) has the form

$$\text{const}\left(\frac{3^{3/2}}{4\pi^3}\right)^{N/2}\Gamma\left(\frac{N+r+M}{2}\right)(-g)^N \quad (9)$$

for even $M$ and, with the extra factor $eN^{1/4}$, for odd $M$ values.

---

[1] The direct expansion of Eq. (2) in powers of the last term in Eq. (5) is incorrect, because the functional integration will then include the configurations for which result (5) is invalid. The calculation should be carried out by the saddle-point method, which yields a continuous function of $N$; the fact that it must be taken at the integer or half-integer points is an external condition.

High-order coefficients in the expansion of the Gell-Mann–Low function $\beta(g) = \sum_N \beta_N(-g)^N$ coincide, except for a constant factor, with the coefficients for the invariant charge [3], which is determined in the electrodynamics by the quantity $gD$, where $D$ is the photon propagator ($M = 2$, $L = 0$). The general asymptotic term is $D_N(-g)^{N+1} \sim NZ_N(-g)^{N+1}$ or $NZ_{N-1}(-g)^N \sim N^{1/2}Z_N(-g)^N$, from whence it follows that

$$\beta_N = \text{const} \times 4.886^{-N}\Gamma\left(\frac{N+12}{2}\right), \quad N \longrightarrow \infty. \quad (10)$$

The same result is obtained if the invariant charge is determined through the triple vertex ($M = 1$, $L = 1$). In this case, the dominant contribution to the asymptotic expression comes from the amputation of the photon line.

**2.** The following four terms of the $\beta$-function expansion in the MOM scheme are known [11]

$$\beta(g) = \frac{4}{3}g^2 + 4g^3 + \left[\frac{64}{3}\zeta(3) - \frac{202}{9}\right]g^4 \\ + \left[186 + \frac{256}{3}\zeta(3) - \frac{1280}{3}\zeta(5)\right]g^5 + \ldots. \quad (11)$$

The series summation procedure should be somewhat modified as compared to [1, 2], because Lipatov's asymptotic expression has the form $ca^N\Gamma(N/2+b)$ instead of $ca^N\Gamma(N+b)$. The Borel transform $B(z)$ is defined as

$$\beta(g) = \int_0^\infty dx e^{-x} x^{b_0-1} B(ag\sqrt{x}),$$

$$B(z) = \sum_{N=0}^\infty B_N(-z)^N, \quad B_N = \frac{\beta_N}{a^N\Gamma(N/2+b_0)}, \quad (12)$$

where $b_0$ is an arbitrary parameter. The conformal mapping $z = u/(1-u)$ of the Borel transform provides a convergent series in $u$ with the coefficients

$$U_N = \sum_{K=1}^N B_K(-1)^K C_{N-1}^{K-1} (N \geq 1), \quad U_0 = B_0, \quad (13)$$

whose large-$N$ behavior

$$U_N = U_\infty N^{\alpha-1}, \quad U_\infty = \frac{\beta_\infty}{a^\alpha \Gamma(\alpha)\Gamma(b_0 + \alpha/2)} \quad (14)$$

determines the parameters of the asymptotic expression $\beta(g) = \beta_\infty g^\alpha$ at $g \longrightarrow \infty$.

The interpolation is performed for the reduced coefficient function

$$F_N = \frac{\beta_N}{\beta_N^{as}} = A_0 + \frac{A_1}{N - \tilde{N}} + \frac{A_2}{(N - \tilde{N})^2} + \ldots, \quad (15)$$
$$\beta_N^{as} = a^N N^{\tilde{b}} \Gamma(N/2 + b - \tilde{b}),$$

by cutting off the series and choosing the coefficients $A_K$ so that Eq. (15) coincides with the known $F_N$ values. Optimal parametrization corresponds to $\tilde{b} = b - 1/2 = 5.5$ [2], while the parameter $\tilde{N}$ is used for checking on the stability of the results and for numerical optimization. In contrast to the $\varphi^4$ theory [1, 2], the common coefficient in the asymptotic expression (10) is unknown. Technically, this is not a problem because the parameter $A_0$ in Eq. (15) is not considered as known but is found by interpolation. However, this leads to a much greater uncertainty in the function $F_N$; its first values (in units of $10^{-3}$) $F_2 = 63.1$, $F_3 = -7.02$, $F_4 = 0.34$, and $F_5 = 1.23$ exhibit only a weak tendency to become a constant, and the predicted value $A_0 = \lim_{N \to \infty} F_N$ changes by several orders of magnitude with changing $\tilde{N}$. At first glance, no reasonable results can be obtained in such a situation.

However, the algorithm used for determining the asymptotic form of $\beta(g)$ is, in a sense, "superstable": the addition of an arbitrary $m$th-order polynomial $P_m(N)$ to $B_N$ does not change the coefficients $U_N$ at $N \geq m + 2$ [2]. This property can be generalized for a wide class of smooth functions: a change in $U_N$ caused by the replacement $B_N \longrightarrow B_N + f(N)$, where $f(N)$ is an integer function with rapidly decreasing Taylor-series coefficients, rapidly decreases with $N$. Thus, smooth errors are immaterial even if they are large. In contrast, the nonsmooth errors lead to a catastrophic effect, which can be used for optimization: if the interpolation procedure is not satisfactory, the behavior of $U_N$ at large $N$ cannot be interpreted in terms of a power-law dependence [2].

To check this argumentation, a test experiment was carried out for the $\varphi^4$ theory. The use of complete information [i.e., coefficients $\beta_2$–$\beta_5$ and parameters $A_0$ and $A_1$ in Eq. (15)] gave $\alpha = 0.96 \pm 0.01$ and $\beta_\infty = 7.4 \pm 0.4$ [2]; the same procedure without the use of $A_0$ and $A_1$ gave $\alpha = 1.02 \pm 0.03$ and $\beta_\infty = 1.7 \pm 0.3$. Taking into account that the uncertainty in the coefficient function (estimated through varying $\tilde{N}$ by ~1 near its optimal value) amounts to few percent in the first case and more than an order of magnitude in the second, one can conclude that such a stability of the results is quite satisfac-tory.[2] Clearly, the results obtained below should only be treated as a zero approximation.

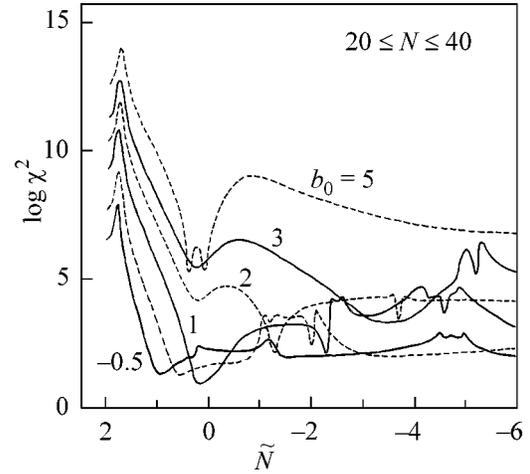

Fig. 1.

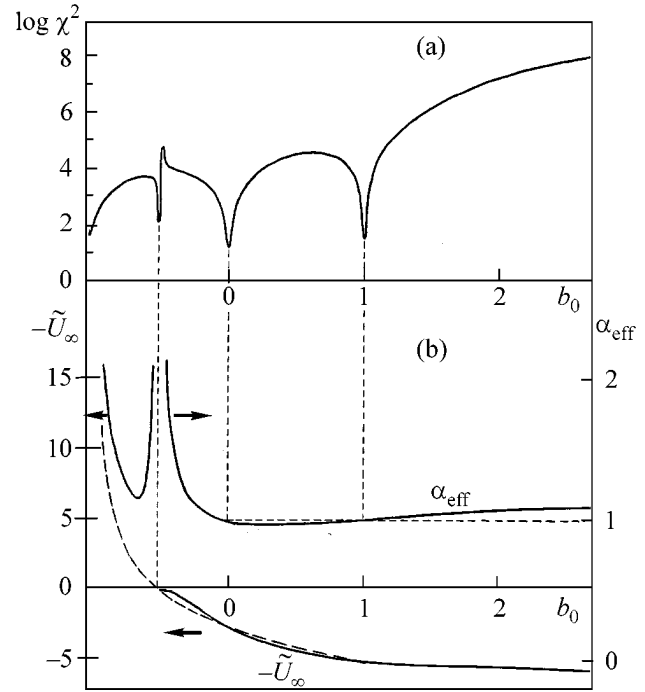

Fig. 2.

Following [2], let us approximate $U_N$ by the power-law dependence for a fixed interval $20 \leq N \leq 40$ and different $b_0$ and $\tilde{N}$ values. The $\chi^2$ dependence on $\tilde{N}$ (Fig. 1) enables one to select a set of interpolations ($-0.5 \lesssim \tilde{N} \lesssim 1.0$) for which the power-law behavior of $U_N$ is probable. The typical dependences of $\chi^2$ and effective values of $U_\infty$ and $\alpha$ on $b_0$ (Fig. 2) indicate that $\alpha \approx 1$.[3] Indeed, the quantity $U_\infty$ reverses its sign [see Eq. (14)] at $b_0 = -\alpha/2 \approx -0.5$. At the same $b_0$ value, $\chi^2$ has a minimum, which corresponds to the fact that the leading contribution $U_\infty N^{\alpha-1}$ vanishes and the power-law dependence $U_N \sim N^{\alpha'-1}$ prevails, where the index $\alpha'$ corresponds to the next correction to the asymptotic expression for $\beta(g)$ (it is assumed that $\beta(g) = \beta_\infty g^\alpha + \beta'_\infty g^{\alpha'} + \beta''_\infty g^{\alpha''} + \ldots$ at large $g$). The values of $\alpha_{\text{eff}}$ at the minima of $\chi^2$ at $b_0 = -\alpha'/2$, $-\alpha''/2$, …, where the respective corrections to Eq. (14) vanish, are closest to the exact value $\alpha \approx 1$ [2].[4]

---

[2] The difference in the $\beta_\infty$ values is not controlled by the estimated error, but this is quite explainable: the procedure proposed in [2] for estimating errors is only justified in the vicinity of the exact result, where all deviations can be linearized.

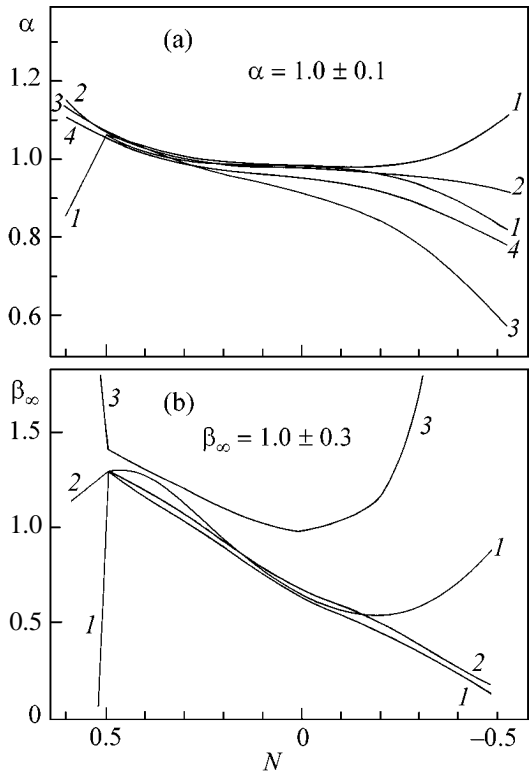

**Fig. 3.**

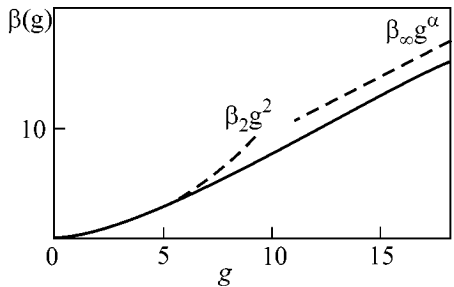

**Fig. 4.**

Figure 3a shows different estimates for the $\alpha$ index as a function of $\tilde{N}$ [2]: (*1*) from the value of $\alpha_{\text{eff}}$ at the $\chi^2$ minima corresponding to $\alpha'$ and $\alpha''$; (*2*) from the position of the $\chi^2$ minimum corresponding to $b_0 = -\alpha/2$; (*3*) from a change in sign of $U_\infty$ when processing by taking the logarithm of $U_N$ (solid line in Fig. 2b); and (*4*) the same but for processing with a fixed index (dashed line in Fig. 2b). Figure 3b shows different estimates obtained for $\beta_\infty$: (*1*) from the $U_\infty$ value at the $\chi^2$ minima corresponding to $\alpha'$ and $\alpha''$ and (*2*) and (*3*) from the slope of the linear portion of the $U_\infty(b_0)$ dependence near the root (upper and lower estimates, respectively). The discrepancy between different estimates gives a measure of uncertainty of the results. For $\tilde{N} \lesssim 0.25$, the results for $\alpha$ are consistent with a value slightly smaller than unity. For $\tilde{N} > 0.25$, there is a systematic increase to 1.08, which is not controlled by the estimated error, but the $\chi^2$ minima are ill-defined and unstable in this case. Similar behavior is observed for $\beta_\infty$. We take, as the most reliable, values in the middle of the chosen $\tilde{N}$ interval, and accept the conservative estimates for the accuracy including systematic changes:

$$\alpha = 1.0 \pm 0.1, \quad \beta_\infty = 1.0 \pm 0.3. \qquad (16)$$

It follows from above that even this estimate of error is not reliable.

It is easy to sum up the series for arbitrary $g$ by calculating the $U_N$ coefficients in Eq. (13) for $N \lesssim 30$ and continuing them according to the asymptotic expression found for $U_\infty N^{\alpha-1}$. Figure 4 shows the results for $\tilde{N} = 0.2$ and $b_0 = 0$. The one-loop law $\beta_2 g^2$ matches the asymptotic dependence $\beta_\infty g^\alpha$ at $g \sim 10$. At $g < 5$, $\beta(g)$ differs only slightly from the one-loop result. Within the accuracy adopted, the asymptotic expression for $\beta(g)$ coincides with the upper limit of inequality $0 \leq \beta(g) < g$, which was derived in [12] from spectral considerations. For $\alpha = 1$ and $\beta_\infty = 1$, the fine structure constant in pure electrodynamics increases at small distances $L$ as $L^{-2}$.

This work was supported by INTAS (grant no. 99-1070) and the Russian Foundation for Basic Research (project no. 00-02-17129).

---

[3] For technical reasons, Fig. 2 shows the quantity $\tilde{U}_\infty = U_\infty \Gamma(b_0 + 1)$.

[4] In the test examples, minima of $\chi^2$ are usually observed only for $\alpha$ and $\alpha'$ [2]. The appearance of additional minima is probably specific to a small amount of information; it was observed in the above-mentioned test experiment for the $\varphi^4$ theory.